\documentclass[twocolumn,superscriptaddress,secnumarabic,amssymb, nobibnotes, aps, bm,pre,floatfix,longbibliography]{revtex4-2}
\usepackage[english]{babel}
\usepackage[latin9]{inputenc}
\usepackage{graphicx}
\usepackage{multirow}
\usepackage[usenames,dvipsnames]{xcolor}
\usepackage{bm}
\usepackage{mathtools}
\usepackage{amsmath}
\usepackage{rotating}
\definecolor{oceanboatblue}{rgb}{0.0, 0.47, 0.75}
\definecolor{orange}{rgb}{1,0.5,0}
\definecolor{goodgreen}{rgb}{0.1,0.5,0}
\definecolor{goodred}{rgb}{0.7,0,0}
\usepackage[colorlinks,urlcolor=goodgreen,citecolor=blue,linkcolor=goodred]{hyperref}
\usepackage{cleveref}
\usepackage{lineno}
\usepackage{tikz}
\usepackage{tocvsec2}
\usepackage{scrextend}
\makeatletter

\newcommand{\comment}[1]{}

\makeatother
\begin{document}
\title{Robust Spin Polarization of Yu-Shiba-Rusinov States in Superconductor/Ferromagnetic 
Insulator Heterostructures}
\author{A. Skurativska}
\email{anastasiia.skurativska@dipc.org}
\affiliation{Donostia International Physics Center (DIPC), 20018 Donostia--San Sebasti\'an, Spain}

\author{J. Ortuzar}
\affiliation{CIC nanoGUNE-BRTA, 20018 Donostia-San Sebastia\'n, Spain}

\author{D. Bercioux}
\affiliation{Donostia International Physics Center (DIPC), 20018 Donostia--San Sebasti\'an, Spain}
\affiliation{IKERBASQUE, Basque Foundation for Science, Plaza Euskadi 5
48009 Bilbao, Spain}

\author{F. S. Bergeret}
\affiliation{Centro de F\'isica de Materiales (CFM-MPC) Centro Mixto CSIC-UPV/EHU,
20018 Donostia-San Sebasti\'an, 
Basque Country, Spain}
\affiliation{Donostia International Physics Center (DIPC), 20018 Donostia--San Sebasti\'an, Spain}

\author{M. A. Cazalilla}
\affiliation{Donostia International Physics Center (DIPC), 20018 Donostia--San Sebasti\'an, Spain}
\affiliation{IKERBASQUE, Basque Foundation for Science, Plaza Euskadi 5
48009 Bilbao, Spain}

\begin{abstract}
Yu-Shiba-Rusinov (YSR) states arise as sub-gap excitations of a magnetic impurity in a superconducting host. 
Taking into account the quantum nature of the impurity spin in a single-site approximation, we study the spectral properties of the YSR excitations of a system of magnetic impurity in a spin-split superconductor, i.e.
a superconductor in proximity to a ferromagnetic insulator at zero external magnetic fields.  
The YSR excitations of this system exhibit a robust spin-polarization 
that is protected from  fluctuations and environmental noise by the exchange field of the ferromagnetic insulator, which can be as large as a few Tesla.  We compare the results of this  quantum approach to the classical approach, which conventionally predicts fully polarized YSR excitations even in the absence of exchange and external magnetic field. Turning on a small magnetic field, we show the latter splits the YSR excitations in the regime where the impurity is strongly coupled to the superconductor, whilst the classical approach predicts no such splitting. 
The studied system can potentially be realized in a tunnel junction connected to a quantum dot in proximity to a spin-split superconductor.  
\end{abstract}
\date{\today}
\maketitle

\section{Introduction}

Magnetic impurities  in superconductors often feature  Yu-Shiba-Rusinov (YSR) excitations. These sub-gap bound states arise due to the exchange coupling between the impurity and the superconductor~\cite{yu,shiba,rusinov}. Much of the recent effort devoted to the study of these excitations is driven by experimental advances in scanning tunneling spectroscopy (STS), which allow to access the spectral properties of YSR excitations with atomic-scale resolution~\cite{eigler:1997, ji:2008, heinrich:2018}.  For example, 
from the spectrum and spatial dependence of the YSR excitations, we can learn about non-conventional pairing properties or the symmetry of the Fermi surface of the host superconductor~\cite{kaladzhyan2016characterizing,ortuzar2022yu}. In addition to magnetic impurities on the surface of superconductors, the YSR excitations have also been investigated in superconducting devices with molecular junctions~\cite{island2017proximity} as well as quantum dots with superconducting leads~\cite{lim2015shiba,jellinggaard2016tuning,valentini2021nontopological}. 

In many theoretical treatments, including the pioneering works of Yu, Shiba, and Rusinov,  magnetic impurities are modeled as classical spins (see e.g.~\cite{Balatsky_2006} for a review). Thus, the impurity is described as an external scattering potential for the quasiparticles of the superconductor. The potential has an opposite sign for opposite spin orientation along the spin-quantization axis,  leading to the two non-degenerate in-gap YSR excitations with opposite energy and full spin polarization. For this reason, systems with the YSR excitations are often proposed as ideal platforms for superconducting spintronics and magnetic characterization at the microscopic scale~\cite{villas2021tunneling,schneider2021atomic,huang2021spin}. However, this description often
overlooks the quantum nature of the spin degree of freedom of magnetic atoms, molecules, or quantum dots~\cite{ruby:2016, choi:2017, machida:2022,franke:2021, schmid:2022,valentini2021nontopological}. Indeed, quantum (and thermal or noise) fluctuations destroy the spin polarization of the YSR excitations. Spin-polarization can be restored by applying external magnetic fields~\cite{wang2021spin,zitko:2017}. However, magnetic fields applied to superconducting devices also have unwanted orbital effects, which may induce supercurrents and suppress superconductivity. 

\begin{figure}[b]
   \centering
    \includegraphics[width=\linewidth]{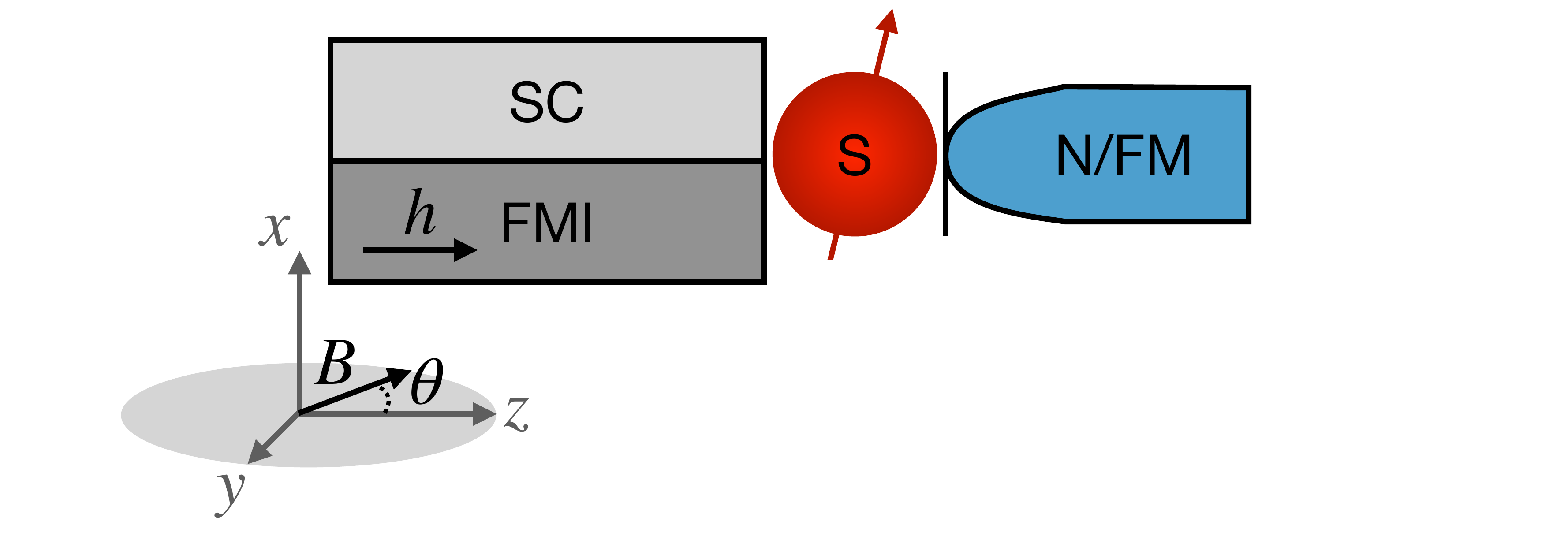}
    \caption{Schematic picture of a possible realization of the studied system. It consists of a superconductor in proximity to a ferromagnetic insulator coupled to a quantum dot or a molecule that can be modeled as a spin-$S$ quantum impurity. The right normal metal is a ferromagnetic tunneling contact to probe the spin polarization. An external magnetic field $\bm B$ is applied in different directions. }
    \label{fig:set-up}
\end{figure}

In this work, we propose using a ferromagnetic insulator (FMI) adjacent to the superconductor to induce a finite spin-polarization of the YSR excitations. 
The FMI leads to an effective exchange field of strength $h$ in the superconductor in the absence of external magnetic field~\cite{Meservey:1991, meservey1994spin,strambini2017revealing,hijano2021coexistence}. The exchange field is induced by the magnetic proximity effect at the FMI/superconductor interface ~\cite{tokuyasu1988proximity,heikkila2019thermal}, and leads to a spin-splitting equivalent to that of a magnetic field as large as tens of Tesla without any orbital effects. To account for quantum fluctuations in the spectrum of a magnetic impurity coupled to such spin-split superconductor, we extend the single-site model used in Refs.~\cite{vecino2003josephson,franke:2021,trivini2022pair}.
This approach provides an excellent qualitative description of the spectra obtained experimentally~\cite{trivini2022pair}. It also captures the properties of the ground state and low-lying states of the system while being computationally cheaper than the numerical renormalization group 
(NRG)~\cite{zitko:2017,bulla:2008}. 
We demonstrate that the exchange field $h$ induced by the FMI suppresses fluctuations and leads to a finite spin-polarization of the YSR excitations without introducing any spin-splitting of the latter.  
In addition, if a small external magnetic field is applied, we show that the YSR excitations split only if the system is in the regime where the impurity spin is strongly coupled to the superconductor.  
In contrast, as we also show below, the classical description of the impurity yields no such spin splitting of the YSR excitations,  the main effect of the external magnetic field being a shift of the energy of the YSR peaks in the spectral function.  

For the sake of simplicity, we focus our analysis on a spin-$\tfrac{1}{2}$ impurity and isotropic exchange coupling between the superconductor and the magnetic impurity. The latter may correspond, for example,  to a quantum dot coupled to an FMI/supeconductor system (see Fig. \ref{fig:set-up}), which can be realized in superconductor/semiconducting nanowire heterostructures in proximity to a FMI \cite{vaitiekenas2021zero}. For other setups relevant to magnetic atoms or molecules on the surface of superconductors, our results can be straightforwardly extended to account for   larger impurity spins, single-ion anisotropy, as well as  anisotropic  exchange~\cite{franke:2021,trivini2022pair}. 
The remaining sections of this article are organized as follows: In the next section, we introduce the model and describe the many-body spectrum of the FMI/superconductor- quantum dot system as a function of the exchange coupling, the exchange field, and the external magnetic field. In Sec.~\ref{section:spectral-properties}, we discuss the spectral properties of the YSR excitations focusing on spin-polarization. Finally, we present our conclusions in Sec.~\ref{section:conclusions}. Appendix~\ref{AppA} contains the details of the classical solution of the model. In Appendix~\ref{AppB}, we provide the details of the analysis of the spin polarization described in Sec.~\ref{section:spectral-properties}.

\section{model and many-body spectrum}

\begin{figure}[t]
    \centering
    \includegraphics[width=\linewidth]{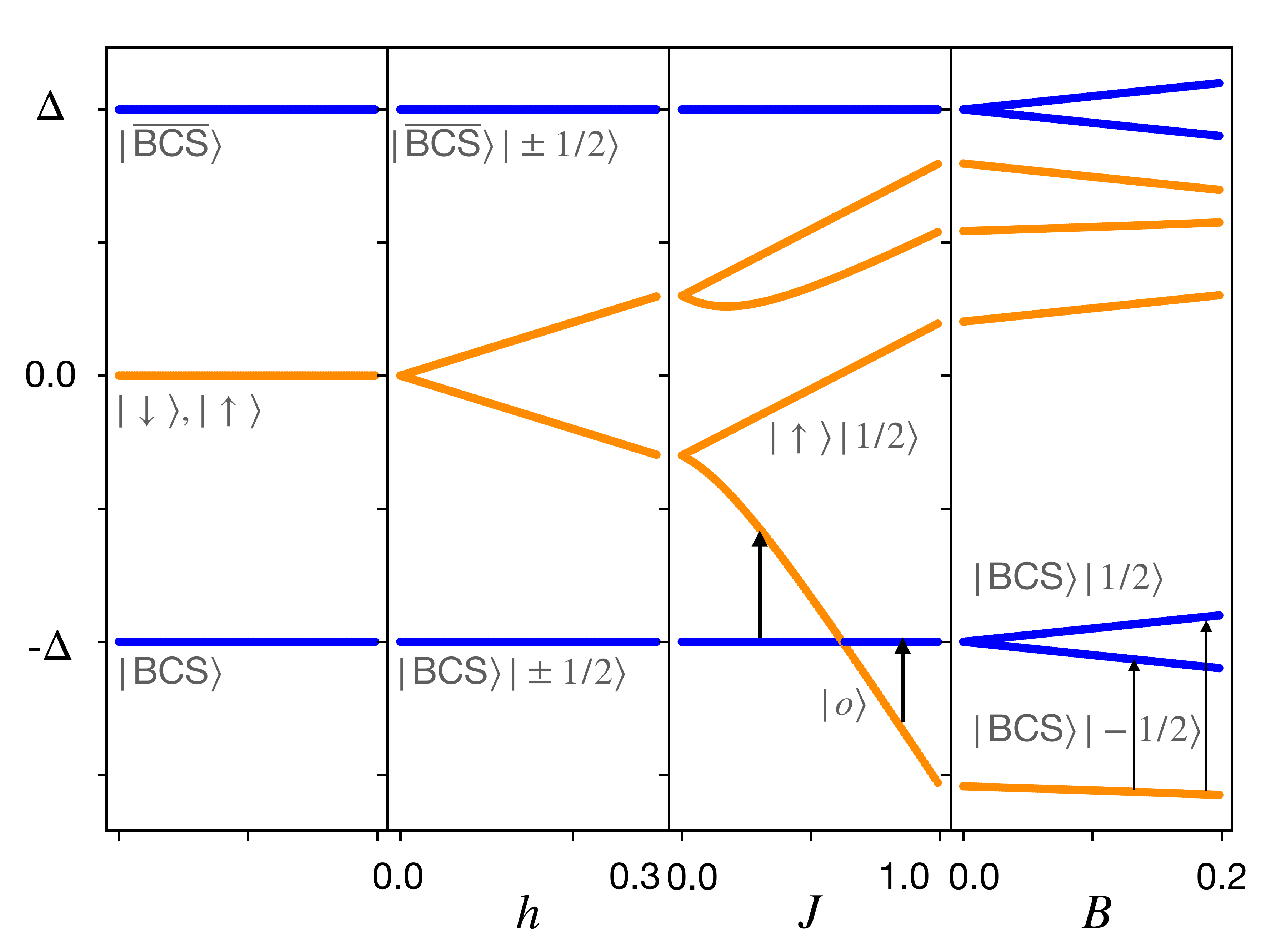}
    \caption{Evolution of the many-body spectrum of the single-site model in Eq.~\eqref{eq:Hamiltonian0} as the exchange field ($h$), exchange coupling ($J$), and external magnetic field ($B$) are switched on sequentially. 
    Their values are given in units of  the strength of the superconducting pairing potential $\Delta$.  The energies of the even and odd fermion-parity states are shown in blue and orange color, respectively. {The expression for the odd-parity eigenstate $|o\rangle$ is given in Eq.~\eqref{eq:oddpgs}.} The arrows indicate the sub-gap transitions corresponding to the YSR excitations  with and without $B$. In the rightmost panel, the system is in the strong coupling regime and the YSR excitations are split by $B$.}
    \label{fig:spectra-phase-diagram}
\end{figure}

 We consider a magnetic impurity in a spin-split superconductor as  schematically shown in Fig.~\ref{fig:set-up}. The exchange field $h$ of the  device is achieved by bringing a conventional ($s$-wave) superconductor in proximity to a ferromagnetic insulator (FMI). Assuming that the thickness of the superconductor is smaller than the superconducting coherence length, it is a good approximation to consider a homogeneous  exchange field $h$ ~\cite{hijano2021coexistence}.  Thus, the  Hamiltonian of the system reads 
 \begin{equation}
\label{eq:Hamiltonian0}
 H = H_0+ H_{J}+ H_{B}\; ,
 \end{equation}
 where
\begin{subequations}\label{eq:many-body-hamiltonian}
\begin{align}
H_0 &= \sum_{{\bm k},\sigma} \xi_{\bm k}c^\dagger_{{\bm k},\sigma}c^{ }_{{\bm k}, \sigma} + \Delta \sum_{\bm k}( c^\dagger_{{\bm k},\uparrow} c^\dagger_{-{\bm k}, \downarrow} + \text{h.c.}) \nonumber\\ & - h \sum_{\bm k}(c^\dagger_{{\bm k}, \uparrow}c^{ }_{{\bm k}, \uparrow} - c^\dagger_{{\bm k}, \downarrow}c^{ }_{{\bm k}, \downarrow})\,, \label{eq:many-body-hamiltonian_1}\\
H_J &= J \sum_{{\bm k}, \sigma \sigma' }c^\dagger_{{\bm k},\sigma}{\bm S} \cdot {\bm s}_{\sigma \sigma'} c^{ }_{{\bm k},\sigma'} \label{eq:many-body-hamiltonian_2} \,,
\\H_B &= {\bm B} \cdot {\bm S} \label{eq:many-body-hamiltonian_3}.
\end{align}
\end{subequations}
Here, $H_0$ describes a superconductor with mean-field pairing potential of strength  $\Delta$ and an exchange field ${\bm h} = h \bm e_z$  along $z$-axis; $H_J$ is the isotropic exchange interaction between the host superconductor and the magnetic impurity  described by the spin operator $\bm S$ with coupling strength $J$. Finally,  $H_B$ accounts for  the Zeeman energy due to  an external magnetic field $\bm{B} = B(\cos \theta, \sin \theta)$ where 
the angle $\theta$  (see Fig.~\ref{fig:set-up}) measures the tilt between the magnetic field and the $z$-axis. 
 
In Eq.~\eqref{eq:many-body-hamiltonian}, the operator $c^\dagger_{{\bm k},\sigma} (c^{ }_{{\bm k},\sigma})$ creates (annihilates) an electron with the momentum $\bm k$, the spin state $\sigma \in \{\uparrow,\downarrow\}$ and the electron dispersion  (measured from the chemical potential) $\xi_{{\bm k}}$; ${\bm s} $ being the Pauli matrices ${\bm s} = (s^x,s^y,s^z)$. We assume an external magnetic field  $|\bm B| \ll |\bm h|$. In this limit, the effect of the magnetic field on the superconductor can be neglected, while its coupling to the impurity-spin persists and can be used as an additional probe into the properties of the YSR excitations, as discussed below. Furthermore,  let us point out that the range of Zeeman couplings analyzed in this work is different from the regime previously studied in Ref.~\cite{zitko:2017} using the NRG, which applies to a different physical situation.

To solve the model  in Eq.~\eqref{eq:Hamiltonian0} we use two different approaches: On the one hand, a single-site model in which the superconducting host is effectively modeled by a single site but its coupling to the impurity spin is described exactly by treating $\bm{S}$ as spin-$\tfrac{1}{2}$ operator. On the other hand,  the classical   description in which the superconductor is treated  as an extended system but the exchange coupling is simplified by treating the impurity spin $\bm{S}$ as a classical vector. 
 
In the single-site model, we  simplify the Hamiltonian $H$ describing the system~\eqref{eq:many-body-hamiltonian}  to the following model:
\begin{subequations}\label{eq:model-hamiltonian}
\begin{align}
    H_0 &= \Delta (c^\dagger_{\uparrow}c^\dagger_{\downarrow} + \text{h.c.}) - h (c^\dagger_{\uparrow}c^{ }_{\uparrow} - c^\dagger_{\downarrow}c^{ }_{\downarrow})\label{eq:model-hamiltonian_a} \,,\\
    H_J &= J \sum_{\sigma \sigma'} c^\dagger_{\sigma} {\bm S}\cdot {\bm s}_{\sigma \sigma'} c_{\sigma'} \label{eq:model-hamiltonian_b}\,,\\
    H_B &= {\bm B}\cdot {\bm S} \label{eq:model-hamiltonian_c}\,.
\end{align}
\end{subequations}
This model is an extension of the single-site model introduced in Ref.~\cite{vecino2003josephson,franke:2021}, which   takes into account the exchange field $h$  due to proximity to the FMI as well as the external magnetic field $\bm B$~\footnote{In contrast to Ref.~\cite{vecino2003josephson},  for  the quantum dot setup of Fig.~\ref{fig:set-up},  we consider here only the Kondo regime in which the dot is singly occupied.}.
Within the single-site model and for a spin-$\tfrac{1}{2}$ impurity, the Hilbert space of the model in Eq.~\eqref{eq:Hamiltonian0} is the tensor product of the four-dimensional Hilbert space of the single superconductor site and  the two-dimensional Hilbert space of the impurity-spin: $\mathcal{H} = \left\{ (|0\rangle ,\, |\uparrow\downarrow\rangle\equiv|2\rangle,\, |\!\uparrow \rangle ,\, |\!\downarrow\rangle ) \otimes ( |\pm 
\tfrac{1}{2}\rangle\right)\}$, where we have defined $|\sigma = \{ \uparrow, \downarrow\} \rangle= c^\dagger_{\sigma} | 0 \rangle$ and $| 0 \rangle$ is the zero-particle state. The Hamiltonian conserves the fermion parity, which for the single-site model takes the form $\mathcal{P} = \prod_\sigma(-1)^{n_\sigma}$,
where $n^{ }_\sigma = c^\dagger_\sigma c^{ }_\sigma$. Thus, all eigenstates can be labeled by their fermion parity and therefore the Hilbert space splits into the direct sum of the even ($\mathcal{P}=+1$) and odd ($\mathcal{P}=-1$) parity sectors, i.e. $\mathcal{H} = \mathcal{H}_\text{e} \oplus \mathcal{H}_\text{o}$ with $\mathcal{H}_\text{e} = \{ (| \text{BCS} \rangle,| \overline{\text{BCS}} \rangle ) \otimes (|\pm \tfrac{1}{2} \rangle ) \}$ and $\mathcal{H}_\text{o}~=~\{ (|\!~\uparrow~\rangle,|~\!\downarrow~\rangle ) \otimes (|\pm 1/2 \rangle ) \}$. Here we have introduced the notation  $|\text{BCS}\rangle = \tfrac{1}{2} (|2\rangle + |0 \rangle )$, $|\overline{ \text{BCS} }\rangle = \tfrac{1}{2} (|2\rangle - |0\rangle)$ for the eigenstates of $H_0$ with eigenvalues $-\Delta$ and $\Delta$, respectively. The single quasi-particle excitations of the superconductor are denoted by $|\!\uparrow\rangle$ and $|\!\downarrow \rangle$ and have zero eigenvalue of $H_0$ at $h=0$. 

Figure~\ref{fig:spectra-phase-diagram} shows the evolution of the many-body spectrum of the system as a function of the  exchange field $h$, the  coupling $J$, and the external magnetic field ${\bm B}  = B \bm {e_z}$ as obtained from the exact diagonalization of the Hamiltonian~\eqref{eq:Hamiltonian0} in the single-site approximation. The leftmost panel shows the spectrum of the Hamiltonian in  
Eq.~\eqref{eq:model-hamiltonian_a} with $h=0$. 

We next discuss the effect of different couplings as we add them sequentially. An exchange field $h$ lifts the degeneracy of the quasi-particle states giving rise to  two two-fold odd-parity degenerate states $|\!\uparrow \rangle| \pm \tfrac{1}{2}\rangle$ and $|\!\downarrow \rangle| \pm \tfrac{1}{2}\rangle$  with energies $\pm h$.  As we show below, this splitting in the presence of the magnetic exchange with the impurity leads to the non-zero polarization of the YSR excitations. The exchange interaction $H_J$ entangles the impurity doublet $|\pm \tfrac{1}{2}\rangle$ with the odd-parity states of the superconductor, resulting in a further splitting of  the many-body states.

  However, in the weak coupling regime, i.e. for small values of $J$ compared to $\Delta$, the ground state of the system is in the even-parity sector, and it is the tensor product of the impurity spin-doublet and the BCS ground-state $|\text{BCS}\rangle |\pm 1/2\rangle$. 
In this regime, the system cannot gain much energy by coupling to the magnetic impurity, and therefore, the electrons in the superconductor remain paired,  leaving the impurity spin unscreened. Thus, the ground state is doubly degenerate, and the total spin projection of the ground state on  $\bm{e_z}$ is  $S^z_T = \pm 1/2$.  We shall refer to this ground state as a doublet and assume that the system is in a mixed state with equal probabilities of the two states of the doublet (this results in zero net polarization of the YSR at $h =0$, as discussed in the following section). Applying a finite magnetic field $B$ selects one of the states of the doublet (or a linear combination thereof) as the absolute ground state and induces a finite spin-polarization, polarization which persists even at $h=0$. However, for weak magnetic fields, we expect the  latter not to be robust to thermal fluctuations and environmental noise. This robustness can be achieved with the help of the exchange field $h$ induced in the superconductor by proximity to an FMI. 

At sufficiently large $J$ (strong coupling regime), the ground state becomes the odd-parity 
state with $S^z_T = 0$ resulting from the entanglement of the impurity doublet and one superconductor (spin-split) quasi-particle excitation, which is given by
\begin{equation} \label{eq:oddpgs}
|o \rangle = \frac{1}{\sqrt{1+ \gamma^2_0}}\left(|\!\downarrow \rangle |+\tfrac{1}{2} \rangle - 
\gamma_0 |\!\uparrow\rangle | - \tfrac{1}{2} \rangle\right)\,,
\end{equation} 
where 
$\gamma_0 = (h+\sqrt{h^2+J^2})/J$.
Although the full spin rotation symmetry is broken by the exchange field induced by the FMI, below we shall 
often refer to this state as the singlet. 

The state $|o \rangle$ becomes the ground state at a critical value of the exchange coupling $J_\text{c} = J_\text{c}(h,\Delta)$,
at which the system undergoes a quantum phase transition (QPT). Across the QPT, the fermion parity ${\cal P}$ of the ground state changes.  Since the tunneling of a single electron (or hole) into  the system  changes the fermion parity, only  excitations between  states of opposite parity are accessible using tunneling probes. In particular, the YSR excitations  are  the lowest lying excitations and correspond to transitions between the ground states in different parity sectors (they are indicated by arrows in Fig.~\ref{fig:spectra-phase-diagram}). In the weak coupling regime ($J < J_\text{c}$), the YSR excitation is a transition from the doublet ground state to the singlet state $| o \rangle$ given in Eq.~\eqref{eq:oddpgs}. On the other hand, in the strong coupling regime ($J > J_\text{c}$), the YSR excitation corresponds to a transition from $|o\rangle$ to the doublet ground states in the even-parity sector. 

When an external magnetic field $\bm B$ is applied, it lifts the two-fold degeneracy of the ground state in the even-parity sector. This results in the splitting of the YSR excitations \emph{only} in the strong coupling regime.  In the weak coupling regime, such splitting does not take place because, as explained above, the magnetic field selects one of the states of the even-parity doublet subspace as the absolute ground state.
As we will discuss in Sec.~\ref{section:spectral-properties}, one can regard the splitting of the YSR excitations as  a consequence of the quantum nature of the impurity spin: Since the tunneling electron (hole) can bring back the superconductor from the singlet state $|o\rangle$ to the $|\text{BCS}\rangle$ state, the impurity spin is left unscreened and  free to precess in the external magnetic field. Note that a classical spin would simply align in the direction of the external magnetic field (see discussion below and Appendix~\ref{AppA}).

\begin{figure*}[!t]
    \centering    \includegraphics[width=0.9\textwidth]{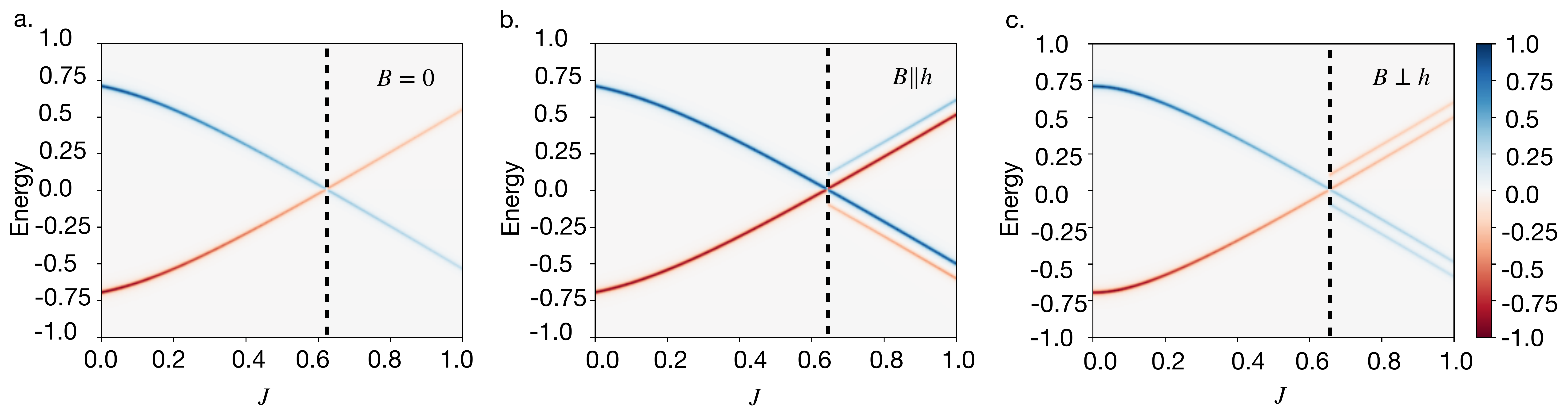}
    \caption{Spin polarization of the YSR excitations as a function of the energy $E$ and the exchange coupling $J$ in the single-site model. In the absence of the magnetic field, the system shows no splitting of the YSR excitations (a), while adding the magnetic field gives rise to the splitting of the YSR excitations in the strong coupling regime (c,d). The dashed line indicates the QPT. All the energies are given in  units of $\Delta$. The values of the parameters used to generate the plots are  $B=0.1$, and $h=0.3$.}
    \label{fig:quantum-model}
\end{figure*}

Although it provides a fully quantum mechanical description of the coupling between the superconductor and the magnetic impurity,  the single-site model described above does not capture many of the effects  of the wide continuum scattering states of the superconductor. Therefore, as far as the spectral properties of the YSR excitations are concerned, the results are rather qualitative and the model is unable to provide information about, e.g., the spatial extent of the excitations. 

Alternatively, the Hamiltonian in Eq.~\eqref{eq:many-body-hamiltonian} can be simplified by modeling the magnetic impurity as a classical spin. Note that, in the case of a quantum impurity, the exchange coupling $H_J$ contains a spin-flip term with non-trivial consequences, especially for impurities with low spin $S$. However, in the classical approach, the impurity spin is treated as a classical vector that aligns with the external magnetic field (when present) and therefore  it can be parametrized as ${\bm S} = S(\cos \theta, \sin \theta )$, where $\theta$ is the angle subtended by the magnetic field $\bm B$ and the exchange field ${\bm h} \propto {\bm e_z}$. This results in a localized spin-dependent scattering potential   proportional to $JS(\cos \theta s_z+ \sin \theta s_x)$ being added to the Bogoliubov-de-Gennes Hamiltonian describing  the superconductor. We refer the reader to Appendix~\ref{AppA}, where we provide further details of the classical approach and describe how the sub-gap spectrum is obtained. In the following section, we will describe the effect of the exchange and  applied magnetic fields on the spectral properties of the YSR excitations and compare the results obtained using the two approaches mentioned 
above. 

\section{Spectral properties of the sub-gap excitations}
\label{section:spectral-properties}

In order to illustrate the consequences of treating  the impurity spin quantum mechanically,  we compare the spectral properties of the YSR excitations in the single-site and classical approaches. Besides the dependence of the excitation  energy on the various system parameters, we are interested in their spin-polarization properties, which can be accessed experimentally using a spin-polarized tunneling probe~\cite{Wiesendanger:2009}. As explained below, the spin polarization of the YSR excitations is defined as the difference of the spectral weight of the spin-up and spin-down YSR peaks of the spectral function measured using a tunneling probe (cf.  Fig.~\ref{fig:set-up}). We normalize the polarization to the maximum of the sum of spectral weights for the two spin orientations of each YSR excitation. 

 Let us briefly recall how the polarization can be measured using a tunneling probe. In the tunneling regime, the full Hamiltonian describing  the tunneling of electrons (or holes) from a tunneling probe contains three terms: 
\begin{equation}
H_{\text{tot}} = H + H_{\text{t}} + H_{\text{ts}}\,, 
\end{equation}
 where $H$ is the system Hamiltonian, which we describe using the single-site model from Eq.~\eqref{eq:model-hamiltonian}, the Hamiltonian for the (spin-polarized) tunneling probe  $H_{\text{t}}$, which is expressed in terms of the creation (annihilation) operators of the electrons in the probe, i.e., $d^{\dagger}_\sigma (d^{}_\sigma)$, and the tunneling Hamiltonian $H_{\text{ts}}$.
For a quantum impurity in the Kondo regime, $H_\text{ts}$ reads (see e.g.~\cite{coleman_2015})
\begin{equation}\label{eq:tunneling Hamiltonian}
H_\text{ts} =  T_0  \sum_{\sigma}c^\dagger_\sigma d_{\sigma}+ T_1 \sum_{\sigma,\sigma^{\prime}}c^\dagger_\sigma {\bm S}\cdot {\bm \sigma}_{\sigma\sigma'} d_{\sigma'}\,,
\end{equation}
where $T_0$ is the direct tunneling amplitude into the superconductor and $T_1$ the tunneling amplitude
through the magnetic impurity, respectively. Notice that the system operators appearing in $T_0$ (e.g. 
$c^{\dag}_{\uparrow}$, for $\sigma = \uparrow$) and $T_1$ (e.g. $c^{\dag}_{\downarrow} S^{+} + c^{\dag}_{\uparrow} S^z$, for $\sigma = \uparrow$) when acting upon a given state invert its fermion parity and change  $S^z_T$ by $\pm \tfrac{1}{2}$. Thus, for zero external magnetic field $B$, the contributions to the  normal current in the weak tunneling regime~\cite{mah00} are of the order $|T_0|^2$, and $|T_1|^2$. Furthermore, when the magnetic field $\bm B$ is not aligned with the exchange field ${\bm h} \propto {\bm e_z}$, $S^z_T$ is not a good quantum number and there is also an interference term proportional $|T^*_0 T_1|\sim B$. However, for the small magnetic fields considered here, we shall neglect this correction. In addition, since the single-site model  only provides a \emph{qualitative} description of the spectral amplitudes, below we focus on the $|T_0|^2$ contribution to the tunneling current only. Indeed,  since the involved operators obey the same selection rules, the $|T_1|^2$ contribution results from transitions between the same many-body states and simply yields an additional (positive) contribution to the current. Focusing
on the $|T_0|^2$ contribution and using the standard tunneling formalism~\cite{mah00}, the spin-polarized tunneling current is determined by the spin-resolved spectral function $A_\sigma(\omega)$, which is obtained from
the imaginary part of the local Green's  function, $A_\sigma(\omega) = -\text{Im} [G^\text{R}_\sigma(\omega)]/\pi$, where $G^\text{R}_\sigma(\omega)$ is the Fourier transform of
\begin{equation}
G^\text{R}_{\sigma}(t) = -i \theta(t) \langle \left\{c_{\sigma}(t), c^{\dag}_{\sigma}(0) \right\}\rangle.  
\end{equation}
Hence, for $\omega > 0$, the spectral function takes the form~\footnote{Since we assume a particle-hole symmetric version of the impurity model (i.e., no scattering potential), the $\omega < 0$ part of $A_{\sigma}(\omega)$ can be simply obtained by the replacement with $\omega\to -\omega$.}:
\begin{equation}
A_\sigma(\omega) = \sum_{n} |\langle \psi_n | c^\dagger_{\sigma} | \psi_0 \rangle |^2 \delta(\omega - \epsilon_n + \epsilon_0) \,.
\end{equation}
Below, we focus on the YSR excitations which correspond to transitions from the ground state of the system,
$|\psi_0\rangle$ to the lowest-lying excited state $|\psi_1\rangle$ (or states for $B \neq 0$ and $J > J_c$, see below). The spectral weight of the YSR excitations is thus given by:
\begin{equation}
Z_{\sigma} = |\langle \psi_1 | c^\dagger_{\sigma} | \psi_0 \rangle |^2\,.
\end{equation}
Hence, we define a (normalized) polarization spectral function for the YSR excitations as follows:
\begin{equation}
\label{def:polarization}
P(\omega) =  \left(\frac{Z_{\uparrow}-Z_{\downarrow}}{Z_{\uparrow}+Z_{\downarrow}} \right) 
\delta(\omega - \epsilon_1 + \epsilon_0) \,,
\end{equation}
where the maximum in the normalization corresponds to the sum of the spectral weights with spin-up and spin-down YSR excitations. 
 For $B\neq 0$, the above expressions must be generalized to include  all the relevant low lying states involved in the YSR excitation (see Fig.~\ref{fig:spectra-phase-diagram}). 
 Further details of the calculations  in the single-site approach are relegated to the Appendix~\ref{AppB}. 
 For the classical approach, the polarization of the YSR excitations is obtained by extending the scattering
 solution of Yu, Shiba, and Rusinov~\cite{yu,shiba,rusinov} to take into account the exchange field $h$, with 
 the details of these calculations being provided in Appendix~\ref{AppA}. Below, we will compare the
 above polarization spectral function to the results of the normalized polarization obtained 
 from the classical approach.

However, before fully discussing the results of those calculations, let us clarify a subtle issue regarding the calculation of the polarization of the YSR excitations in the single-site approach.  Let us recall that, in the weak coupling regime at zero magnetic field, the ground state of the system is  the doublet  $|e_{\pm} \rangle = |\text{BCS}\rangle |\pm \tfrac{1}{2}\rangle \,$. An unbiased preparation of the system will result in the ground state being either $|e_+\rangle$ or $|e_-\rangle$ with  equal probability, which is described by the following mixed state:
\begin{equation}
\label{eq:mixed-state}
\rho_e = \frac{1}{2} \left(|e_+ \rangle \langle e_+ |+ 
|e_- \rangle \langle e_- | \right). 
\end{equation}
In this expression, the pre-factor $p_{i=\{\pm\}} = \frac{1}{2}$ refers to the classical probability for the system to be found in one of the states of the doublet. Therefore, the expression for the spectral function needs to be modified in order to take into account that the ground state is a mixed state, which results in the following expression:
\begin{equation}
\label{eq:spectral-fn-weak-coupling}
A_\sigma(\omega) = \sum_{i=\{\pm\}}p_i \sum_{n} |\langle \psi_n | c^\dagger_{\sigma} | e_i \rangle |^2 \delta(\omega - \epsilon_n + \epsilon_0)\,. 
\end{equation}
Note that, in the absence of the exchange and external magnetic fields (i.e. $h = B = 0$), a tunneling electron (hole) will induce a transition to a state that has a non-zero overlap with the lowest energy odd-parity state,
\begin{equation}
|o\rangle = \frac{1}{\sqrt{2}}(|\!\downarrow\rangle|+\tfrac{1}{2}\rangle-|\!\uparrow\rangle|-\tfrac{1}{2}\rangle)\,. \label{eq:oddgsh0b0}
\end{equation}
This yields equal spectral weight of the YSR excitation for the two spin orientations, i.e. $Z_\uparrow = Z_{\downarrow} = \tfrac{1}{16}$, hence resulting in zero spin-polarization. Zero polarization is also obtained when the calculation is carried out in the strong coupling regime, in which the ground state is a pure state corresponding to the odd-parity singlet $|o\rangle$ from Eq.~\eqref{eq:oddgsh0b0}.  On the other hand,  in the classical approach in the absence of $h$ and $B$, the classical vector describing the spin of the magnetic impurity is \emph{conventionally} chosen along a certain direction (the spin quantization axis).  Thus,  the solutions of the BdG equations, including the YSR in-gap levels, have the spin projection on the spin quantization axis as a good quantum number. This has led to the perception that the YSR excitations are indeed spin polarized in both weak and strong coupling regimes. Below, when considering the classical approach,  we shall follow the same convention.  

\begin{figure}[!b]
    \centering
    \includegraphics[width=\linewidth]{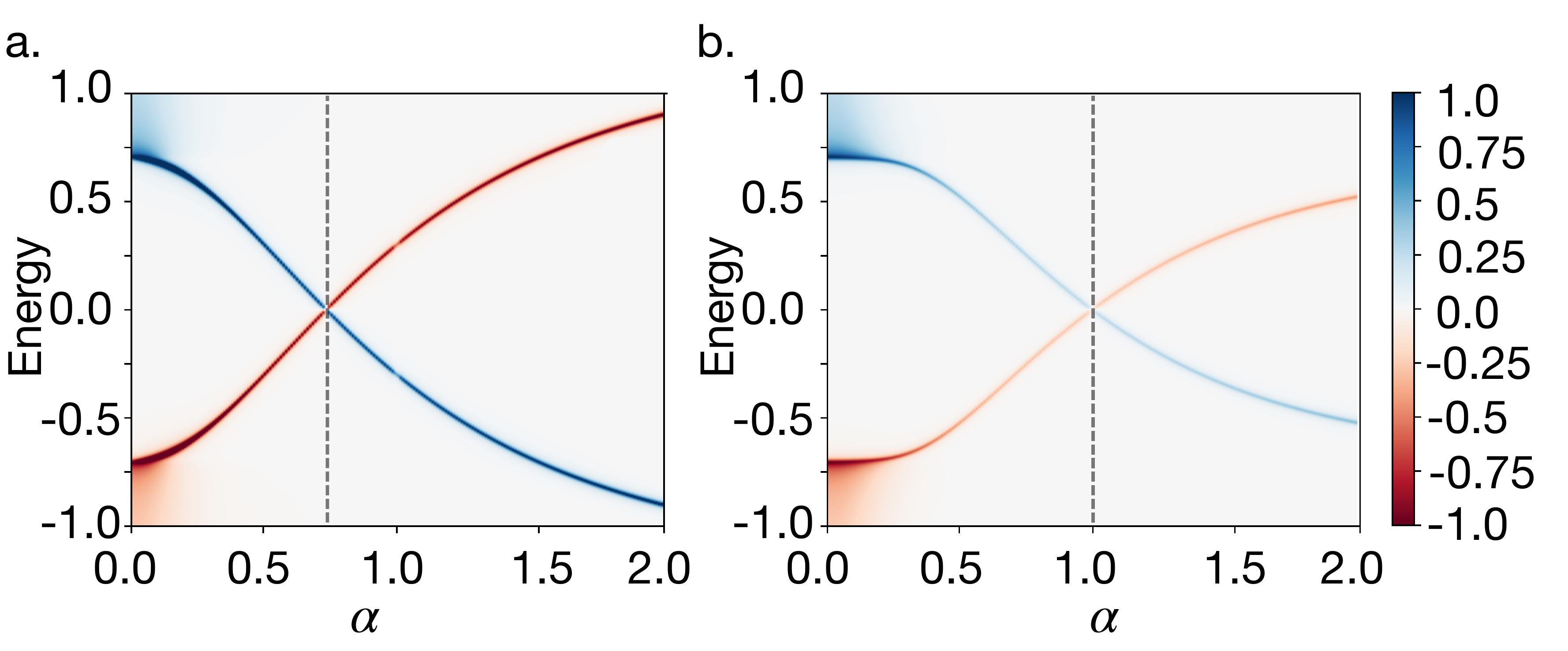}
    \caption{Polarization of the YSR excitations from the classical approach as a function of the dimensionless coupling parameter $\alpha \simeq JS$ for two different impurity-spin $\bm S$ orientations: parallel (a) and perpendicular (b) to the exchange field $h$. In order to highlight the polarization of the YSR excitations here we plot $A_\uparrow(\omega)-A_\downarrow(\omega)$ normalized by the maximum value of the total density of states, that eliminates contribution  from the continuum of states. The dashed line indicates the phase transition. The energies are in units of the superconducting pairing potential $\Delta$. In the two panels, we have set $h=0.3$. }
    \label{fig:classical-model}
\end{figure}

Next, we discuss the polarization function $P(\omega)$ of the YSR excitations as a function of the exchange coupling and the applied magnetic field in the single-site model and compare the results to the classical approach. 
As anticipated above,  we will show that the presence of either the exchange or magnetic field is required for the YSR excitations to have non-zero polarization. In the absence of a magnetic field, the polarization of the states is protected by the exchange field $h$, which can be of the order of a few Tesla, thus, making the polarization robust against thermal fluctuations and environmental noise. Figure~\ref{fig:quantum-model} shows the polarization spectral function $P(\omega)$ of the YSR excitations as a function of the exchange coupling $J$ for the finite value of the exchange field: In the absence of an external magnetic field, the single-site model predicts the existence of a pair of spin-polarized YSR excitations both in the weak and in the strong coupling limit (see Fig.~\ref{fig:quantum-model}(a)). The polarization spectrum in the classical model shows a somewhat similar behaviour: two spin-polarized YSR excitations crossing at the critical value of the exchange coupling (see Fig.~\ref{fig:classical-model}(a)). However, closer examination reveals a crucial difference between the two approaches: While in the classical limit, the YSR excitations are fully polarized for any value of $J$ (cf.  the Fig.~\ref{fig:classical-model}), in the quantum approach, the polarization of the YSR excitations at $\omega > 0$ depends on the exchange field $h$ and coupling $J$ as follows: 
\begin{equation}
\frac{Z_{\uparrow} - Z_{\downarrow}}{Z_{\uparrow} + Z_{\downarrow}} \propto \frac{\gamma^2_0-1}{\gamma^2_0+1} = \frac{h}{\sqrt{h^2+J^2}}. 
\end{equation}
Thus, at a finite value of $h$, the YSR excitations are polarized even for $B=0$. Applying an external magnetic field  alters the polarization of the YSR excitations. In the presence of the magnetic field, one of the  ground states in the weak coupling regime (or a linear combination of them) is selected and the system is no longer described by a mixed state, which further enhances the polarization. 

\begin{figure}[!b]
    \centering
    \includegraphics[width=\linewidth]{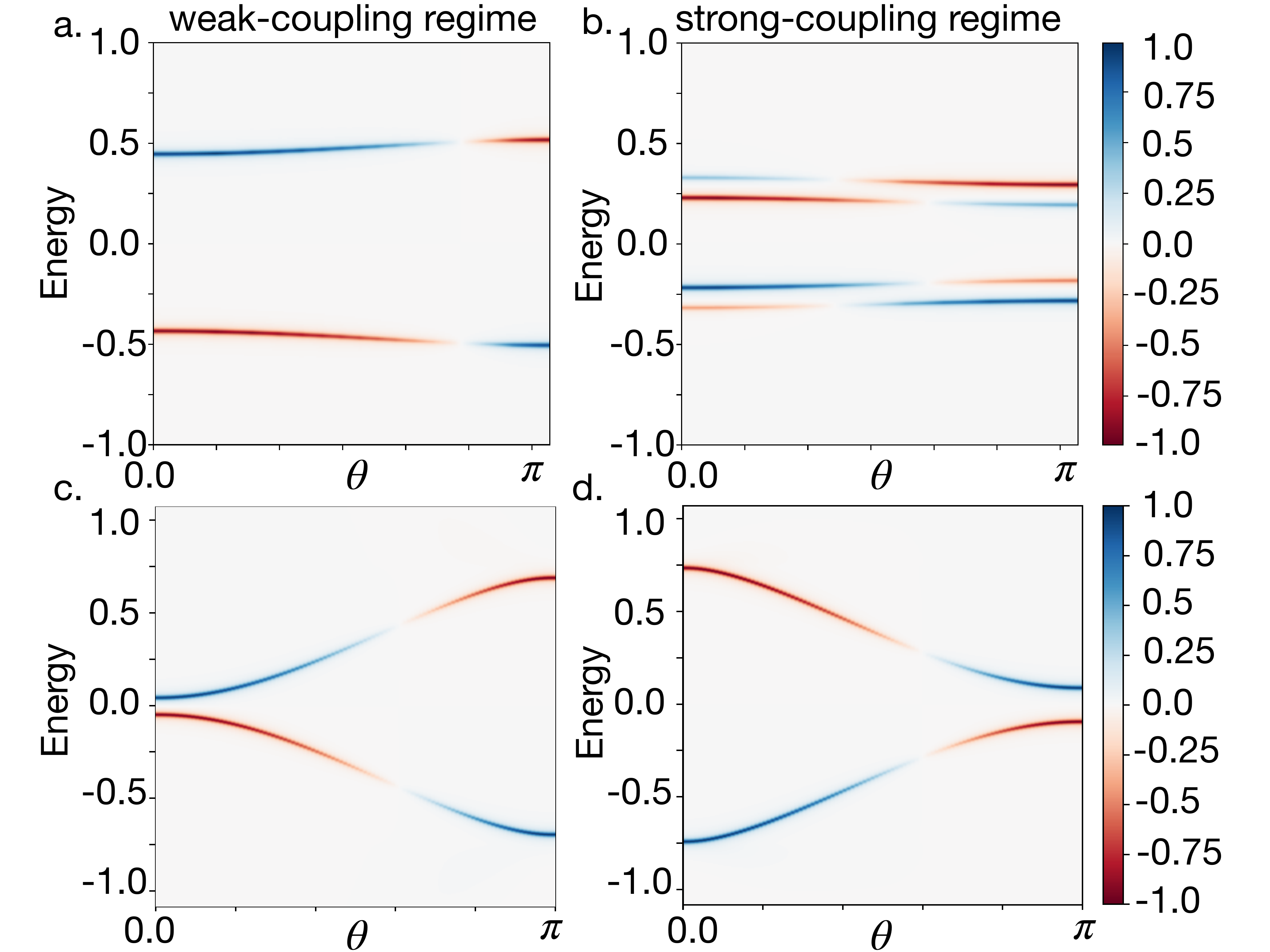}
    \caption{Polarization spectral density as a function of the energy $E$ and the angle of the external magnetic field ${\bm B} = B (\cos \theta, \sin \theta )$ in the weak and strong coupling regime for the single-site (a, b) and classical (c, d) approaches.  Energies are given  in units of  $\Delta$. The choice of system parameters is $B =0.1$, $h=0.3$. }
    \label{fig:angle-dependence}
\end{figure}

Regarding the effects of the external magnetic field, we first consider the case when ${\bm B} || {\bm h}$ in   Fig.~\ref{fig:quantum-model}(b). In the weak coupling regime, the system exhibits a pair of fully-polarized YSR excitations and a similar result is obtained using the classical approach. However,  in the strong coupling regime applying the magnetic field  splits the YSR excitations into two pairs of the sub-gap excitations: the main YSR state and its `satellite' with lower polarization (see Fig.~\ref{fig:quantum-model}(b) for 
$J > J_{\text{c}}$). For ${\bm B} || {\bm h}$ the polarization of the satellite state behaves as $P(\omega) \propto -\gamma^2$, where $\gamma = \left(B+2h+\sqrt{(B + 2h)^2+4J^2}\right)/2J$, and it has the opposite sign compared to the main YSR state spin-orientation. For the details of the polarization calculations see Appendix~\ref{AppB}. 

Changing the orientation of the external magnetic field allows to control the polarization of the pair of sub-gap excitations as shown in Fig.~\ref{fig:quantum-model}(c): By applying the magnetic field perpendicularly to the direction of the exchange field  reverses the spin polarization of the satellite peaks, such that two states have the same polarization orientation. For ${\bm B} \perp {\bm h}$ the polarization of both the main YSR peaks and their satellites decreases with increasing exchange coupling $J$.

 To summarize, an external magnetic field can  be used as a knob for tuning the spin-polarization of the YSR excitations. Figure~\ref{fig:angle-dependence} shows the polarization of the YSR excitations as a function of the angle subtended by the applied magnetic and exchange fields $\theta \in (0,\pi )$. In the single-site approach, the behaviour of polarization depends on the strength of the exchange coupling: in the weak coupling regime, the spin polarization of the YSR excitations switches with $\theta$, while in the strong coupling regime, due to the splitting of the YSR excitations in the magnetic field, the response is qualitatively different: the switching happens between the main YSR state and its satellite. Additionally, the switching of polarization occurs at different values of $\theta$ for the YSR state and its satellite. Thus, there is a range of $\theta$ around $\pi/2$ for which there is a pair states with the same finite spin polarization. On the other hand, as mentioned above, in the classical case, there is no qualitative difference in the polarization behaviour between the strong and weak coupling regimes.

\section{Conclusions}~\label{section:conclusions}

We have studied a system consisting of a ferromagnetic insulator/superconductor structure coupled to a quantum dot in the Kondo regime. 
We demonstrate that the spin-splitting induced in the superconductor via the magnetic proximity effect leads to spin polarization of the YSR excitations even in the absence of an external magnetic field. 

To capture the quantum nature of the quantum dot spin in a qualitative fashion, we employed a single-site model describing a quantum impurity coupled to a spin-split single-site superconductor.
This model, despite its simplicity,  correctly captures the many-body nature of the system's ground state, in particular, the QPT occurring as a function of the exchange coupling, accompanied by the change in the fermion parity and the total spin of the ground state. Both the weak and the strong coupling phases are characterized by the low-energy spin-polarized YSR excitations.

We find that the single-site model predicts the splitting of the YSR excitations in the strong coupling regime, while the classical impurity limit does not describe this splitting. Changing the orientation of the magnetic field allows controlling the polarization of the YSR excitations, namely rotation of the magnetic field allows to switch the polarization of the excitations. 

For applications in spintronics and transport in quantum devices, the main advantage of using an FMI  is that the polarization of the YSR excitations occurs without the need of applying a large external magnetic field which would inevitably affect superconductivity.
Our results can be straightforwardly extended 
to other setups, as for example 
molecules on the surface of superconductors with larger spin number, magnetic anisotropy, as well as  anisotropic exchange coupling~\cite{franke:2021,trivini2022pair}.

\section*{Acknowledgements}

A.S., D.B., and M.A.C. acknowledge the support from the Spanish MICINN-AEI through Project No.~PID2020-120614GB-I00  and the funding from the Basque Government's IKUR initiative on Quantum technologies (Department of Education). F.S.B. acknowledges the support from the Spanish MICINN-AEI through Projects No.~PID2020-114252GB-I00 (SPIRIT) and TED2021-130292B-C42,  the Basque Government through grant IT-1591-22, and the EU's Horizon 2020 Research and Innovation Program under Grant Agreement No.~800923 (SUPERTED). D.B. acknowledges the Transnational Common Laboratory $Quantum-ChemPhys$.

\section*{Appendices}

\appendix

\section{Classical Approach}\label{AppA}
Following Yu, Shiba and Rusinov \cite{yu,shiba,rusinov} original works, the problem of a classical impurity on a superconductor can be analytically solved by accounting for the exchange field in  the bare superconductor Green's function (GF). To this end, notice that the Bogolyubov-de-Gennes (BdG) Hamiltonian describing a spin-split superconductor written in the Nambu basis, i.e. $\Psi_{\bm k}=( c^{ }_{{\bm k} \uparrow}\, c^{ }_{{\bm k}\downarrow}\, c^\dagger_{-{\bm k} \downarrow}\, -c^\dagger_{-{\bm k}\uparrow})^\text{T}$, takes the form:
\begin{equation}
\begin{split}
H &= \sum_{\bm k} \Psi^\dagger_{\bm k} H^{\text{BdG}}_{\bm k}\Psi_{\bm k}\,,\\
H^{\text{BdG}}_{\bm k} &= \epsilon_{\bm k} \tau_3 + \Delta \tau_1 + h \sigma_3 \tau_0\,,
\end{split}
\end{equation}
where $\tau_{i=1,2,3}$ and $\sigma_{i=1,2,3}$ are the Pauli matrices corresponding to the particle-hole and the spin degrees of freedom, respectively. Hence, the unperturbed GF of a spin-split superconductor reads 
\begin{equation}
\begin{split}
\hat G_0^{-1} ( \omega, {\bm k}) &=   i\omega \sigma_0 \tau_0 - H^{\text{BdG}}_{\bm k} \,,\\
\hat G_0(\omega,{\bm k}) &= \frac{ (h-\omega)\tau_0 - \xi_{\bm k} \tau_3 -\Delta \tau_1}{\Delta^2+\xi_{\bm k}-(h-\omega)^2}\,.
\end{split}
\end{equation}
Performing summation over the momenta, we obtain the local GF
\begin{equation}
\hat G_0(\omega) = -\pi \nu \frac{(h-\omega)\tau_0 -\Delta \tau_1}{\sqrt{\Delta^2 - (\omega-h)^2}}\,,
\end{equation}
where $\nu$ is the electron density of states at the Fermi level. 
The exchange coupling in the limit of classical impurity is given by a scattering potential 
$\hat{V}= \frac{J}{2}{\bm{S}}\cdot {\bm{\sigma}} \,,$
where $J$ is an exchange coupling between the impurity-spin $S$ and the spin-density of a superconductor. Note that in the classical limit $\bm{S}$ is a vector. 

We compute the $T$-matrix, whose  poles are the energies of the the sub-gap bound states.
The  $T$-matrix can be defined using the following equation for the perturbed local GF matrix:
\begin{equation}
\hat G ( \omega) = \hat G_0( \omega) + \hat G_0( \omega) \hat T(\omega) \hat G_0 ( \omega) \, . 
\end{equation}
Upon comparing this equation with the Dyson equation, we arrive at $\hat T(\omega) = \hat{V}\left[1-\hat{G}_0(\omega)\hat V \right]^{-1}$. Hence, for an impurity aligned with the external magnetic field, we obtain:f
\begin{equation}
\hat G (\omega) = -\frac{\pi\nu}{D}
\begin{pmatrix}
a & \Delta\\
\Delta & a
\end{pmatrix}\,,
\end{equation}
with  $D = 2\alpha(h-\omega) + (\alpha^2 - 1)\sqrt{\Delta^2 -(h-\omega)^2}$ and $a = h-\omega -\alpha\sqrt{\Delta^2 - (h-\omega)^2}$, where we have introduced the dimensionless parameter 
$\alpha = \pi \nu J S/2$. The local retarded GF $G^{R}(\omega)$ is obtained by replacing
$\omega \to \omega + i \delta$ in the above expression, where $\delta\to 0^{+}$. The spin-resolved spectral function  is  obtained from normal components of the GF matrix using
\begin{equation}
A_{\sigma=\{\uparrow, \downarrow\}}^{\text{cl.}}(\omega) = -\frac{1}{\pi} \text{Im}[G^R_{\sigma\sigma}(\omega)]\,. 
\end{equation}
For $h>0$ $A_\uparrow^{\text{cl.}}(\omega)$ has the YSR peak at $\omega_\uparrow = h - \Delta \frac{ (1 - \alpha^2)}{(1 + \alpha^2)}$, while $A_\downarrow^{\text{cl.}}(\omega)$ has a peak at $\omega_\downarrow = -h + \Delta \frac{ (1 - \alpha^2)}{(1 + \alpha^2)}$. Notice that when the external magnetic and exchange fields are aligned this approach yields two fully spin-polarized YSR excitations. Thus, the exchange field merely introduces a shift of the the YSR peak energy. A closed analytical  expression of the energy of the YSR peaks can also be obtained for $\bm B$  perpendicular to the exchange field $\bm h$, but not in the general case. However, by obtaining the spin polarization numerically we observe that the main difference between the aligned and non-aligned cases is the change in the spin polarization of the YSR excitations, which changes from being fully polarized to partially polarized as the angle $\theta$ between the magnetic and exchange field increases. 

\section{Spin polarization of the YSR excitations in the single-site model}\label{AppB}

In this appendix, we calculate spin polarization of the YSR excitations in the single-site approximation. Assuming non-zero exchange field we obtain the polarization of the YSR excitations analytically for the cases of $B=0$ and ${\bm B} || {\bm h}$. 
The results are shown in Fig.~\ref{fig:quantum-model}(a) and Fig.~\ref{fig:quantum-model}(b).

\begin{figure}
    \centering
    \includegraphics[width=\linewidth]{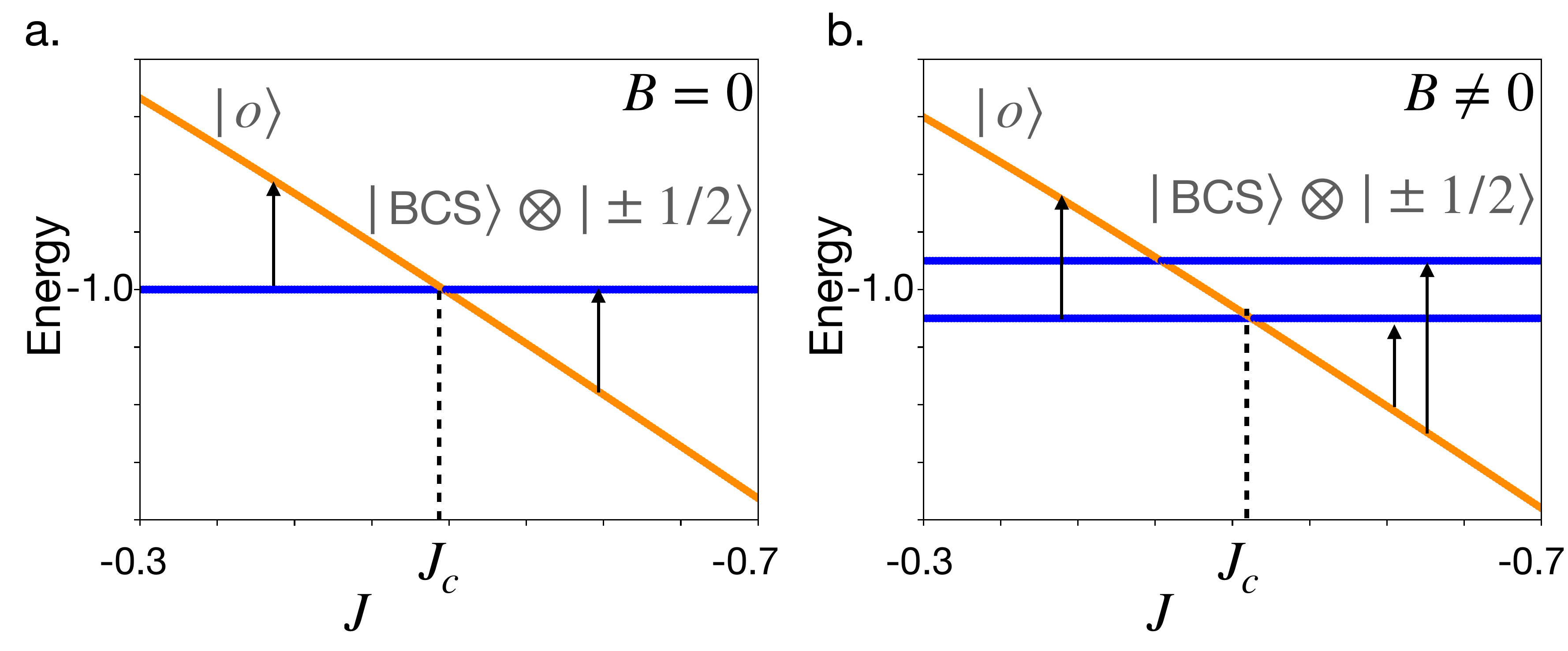}
    \caption{Schematic representation of the low-energy spectrum of the single-site model as a function of exchange coupling $J$ for the case of zero (a) and non-zero (b) magnetic field $B$. Arrows indicate possible transitions in the weak coupling ($J<J_\text{c}$) and in the strong coupling ($J<J_\text{c}$) regimes. }
    \label{fig:low-energy-spectra}
\end{figure}

When $h \neq 0$ and $B = 0$ the polarization $P(\omega)$ is computed using the expression given in Eq.~\eqref{def:polarization} of the main text. 
The low-energy spectrum for this choice of parameters is shown in Fig.~\ref{fig:low-energy-spectra}(a). In the weak coupling regime ($J<J_c$) the ground-state at $\epsilon_0 = -\Delta$ is two-fold degenerate and is described by the density matrix $\rho_e$ in Eq.~\eqref{eq:mixed-state}.
The first excited state with the energy 
$\epsilon_1 = -\frac{(h+ \sqrt{h^2 +J^2})}{J}$ is given by 
\begin{equation}
\label{eq:parity-odd}
| o \rangle = \frac{1}{\sqrt{1+\gamma^2_0}}\left( |\downarrow, 1/2 \rangle  -\gamma_0 |\uparrow, -1/2\rangle \right) \,, 
\end{equation}
where $\gamma_0 = \frac{h + \sqrt{h^2+J^2}}{J}\,.$
The amplitudes of the spectral function are
\begin{equation}
\begin{split}
Z_\uparrow = |\langle o | c^\dagger_\uparrow |\rho_e\rangle |^2 =\frac{\gamma^2_0}{8(1 + \gamma^2_0)}\,,\\ 
Z_\downarrow = | \langle o | c^\dagger_\downarrow |\rho_e\rangle |^2 = \frac{1}{8(1 + \gamma^2_0)}\,.
\end{split}
\end{equation}
In this case, there is a single spin-polarized YSR excitation at $\omega = \epsilon_1-\epsilon_0$ and its polarization is given by
\begin{equation}
P(\omega) = \left( \frac{\gamma^2_0-1}{1 + \gamma^2_0} \right) \delta(\omega -\epsilon_1+\epsilon_0)\,,
\label{eq:pol-zero-field}
\end{equation}
with the amplitude decreasing as a function of the exchange coupling as $\frac{\gamma^2_0-1}{1 + \gamma^2_0} \propto \frac{h}{\sqrt{h^2+J^2}}$ for a fixed value of the exchange field.

When $h \neq 0$ and $\bm B || \bm h$ low-energy states involved in the YSR excitations are shown in Fig.~\ref{fig:low-energy-spectra}(b). Let us discuss the weak and strong coupling regimes separately. For $J<J_c$ the magnetic field selects (with probability $p_-=1$) one of the doublet parity-even states $|e_-\rangle$ as the absolute ground state. The first excited parity-odd state $|o\rangle$ is as in Eq.~\eqref{eq:parity-odd} but with $\gamma = \frac{\left(B+2h+\sqrt{(B + 2h)^2+4J^2}\right)}{2J}$. Hence, the spectral weights for spin-excitations are the following
\begin{equation}
\begin{split}
Z_\uparrow &= |\langle o | c^\dagger_\uparrow |e_-\rangle|^2 = \frac{\gamma^2}{2(1+\gamma^2)}\,,\\ 
Z_\downarrow &= |\langle o | c^\dagger_\downarrow |e_-\rangle|^2 = 0\,.
\end{split}
\end{equation}
Therefore, in the weak coupling regime, there is a single YSR peak with constant polarization intensity $P(\omega) \propto (Z_\uparrow-Z_\downarrow)/ (Z_\uparrow+Z_\downarrow) = 1$. 
The strong coupling regime requires more care. For $J>J_c$ the ground state is given by the odd-parity singlet $|o\rangle$ and there are two even-parity states $|e_\pm\rangle$ the electron can tunnel to. The polarization in this case has two contributions

\begin{equation}
P(\omega) = \frac{\sum_{i=\{ \pm \}} (Z^i_\uparrow-Z^i_\downarrow)\delta(\omega - \epsilon_i+\epsilon_0)}{\max[Z^+_\uparrow+Z^+_\downarrow,Z^-_\uparrow+Z^-_\downarrow]}\,,
\end{equation}
where $\epsilon_{i=\{\pm\}} = -\Delta \pm \frac{B}{2}$. The spectral weights are
\begin{equation}
\begin{split}
Z^{-}_\uparrow &= |\langle e_-|c^\dagger_\uparrow | o \rangle|^2 = 0\,,\\
Z^{-}_\downarrow &= |\langle e_-|c^\dagger_\downarrow | o\rangle|^2 = \frac{\gamma^2}{2(1 + \gamma^2)}\,,\\
Z^{+}_\uparrow &= |\langle e_+|c^\dagger_\uparrow | o\rangle|^2 = \frac{1}{2(1 + \gamma^2)}\,,\\
Z^{+}_\downarrow &= |\langle e_+|c^\dagger_\downarrow | o\rangle|^2 = 0\,.
\end{split}
\end{equation}
The polarization is given by the sum $P(\omega) = P_+(\omega)+P_-(\omega)$ with
 \begin{equation}
 \begin{split}
P_+(\omega) &= \frac{Z^{+}_\uparrow}{Z^{+}_\uparrow}\delta(\omega - \epsilon_+ +\epsilon_0) \,,\\
P_-(\omega) &= -\frac{Z^{-}_\downarrow}{Z^{+}_\uparrow}\delta(\omega - \epsilon_- +\epsilon_0) \,,
  \end{split}
 \end{equation}
where we normalize each term by $Z^+_\uparrow$, because $Z^+_\uparrow~>~Z^-_\downarrow$ for $J>J_c$. 
Two components $P_\pm(\omega)$ correspond to the main YSR excitation and its satellite, respectively. These two states show different behaviour as a function of the exchange coupling $J$: the polarization intensity of the satellite state $P_- \propto -\gamma^2$, with $\gamma = \frac{B+2h}{\sqrt{(B+2h)^2+4J^2}}$ increases as a function of $J$, while the polarization of the main YSR excitation $P_+(\omega)$ stays constant as a function of $J$ and has an opposite spin-orientation.

\bibliography{biblio}

\end{document}